\documentclass[10pt,a4paper]{article}

\usepackage{times}

\usepackage{epsfig,wrapfig}
\usepackage{epstopdf} 
\usepackage{subfigure,color}

\usepackage{fancyhdr}

\usepackage[lmargin=2.5cm, rmargin=2.5cm,tmargin=3.50cm,bmargin=3.50cm]{geometry}
\usepackage{indentfirst}

\usepackage{titlesec}
\titleformat{\section}[hang]
  {\centering}{\thesection}{1ex}{\normalsize \textsc}
\titleformat{\subsection}[hang]
  {}{\thesubsection}{1ex}{\normalsize \textit}

\newcommand{\acknowledgement}{\section*{\centering{\textnormal{\normalsize{\textsc{Acknowledgement}}}}}}

\renewcommand{\thesection}{ \normalsize \textnormal{\Roman{section}.}}
\renewcommand{\thesubsection}{\normalsize \textnormal{\textsc{\textit{\Alph{subsection}.}}}}

\def\e{\begin{equation}}
\def\f{\end{equation}}
\def\_#1{{\bf #1}}

\def\E{\varepsilon}

\def\.{\cdot}

\begin{document}

\title{\large \textbf{Electromagnetic Aspects of Practical Approaches to Realization of Intelligent Metasurfaces}}
%
\def\affil#1{\begin{itemize} \item[] #1 \end{itemize}}
\author{\normalsize \bfseries \underline{F.~Liu}$^1$, \underline{O.~Tsilipakos}$^2$, X.~Wang$^1$, A.~Pitilakis$^{2,3}$, A.~C.~Tasolamprou$^2$, M.~S.~Mirmoosa$^1$, D.-H.~Kwon$^4$\\ \normalsize \bfseries K.~Kossifos$^{5}$, J.~Georgiou$^{5}$, M.~Kafesaki$^{2,6}$, C.~M.~Soukoulis$^{2,7}$ and S.~A.~Tretyakov$^1$
}
\date{}
\maketitle
\thispagestyle{fancy} 
\vspace{-6ex}
\affil{\begin{center}\normalsize $^1$Aalto University, Department of Electronics and Nanoengineering, Maarintie 8, 02150, Espoo, Finland\\
$^2$Foundation for Research and Technology Hellas, Heraklion, 71110, Crete, Greece\\
$^3$Aristotle University of Thessaloniki, Dept. of Electrical and Computer Engineering, Thessaloniki, Greece\\
$^4$University of Massachusetts Amherst, Dept. of Electrical and Computer Engineering, Massachusetts, USA\\
$^5$University of Cyprus, Department of Electrical and Computer Engineering, 1 Panepistimiou Avenue, 2109 Aglanzia, 1678 Nicosia, Cyprus\\
$^6$University of Crete, Department of Materials Science and Technology, Heraklion, 71003, Crete, Greece\\
$^7$Ames Laboratory and Dept. of Physics and Astronomy, Iowa State University, Ames, Iowa 50011, USA\\
fu.liu@aalto.fi; otsilipakos@iesl.forth.gr
\end{center}}

\begin{abstract}
\noindent \normalsize
\textbf{\textit{Abstract} \ \ -- \ \
We thoroughly investigate the electromagnetic response of intelligent functional metasurfaces. We study two distinct designs operating at different frequency regimes, namely, a switch-fabric-based design for GHz frequencies and a graphene-based approach for THz band, and discuss the respective practical design considerations. The performance for tunable perfect absorption applications is assessed in both cases.}
\end{abstract}

\section{Introduction}

Metasurfaces are artificial two-dimensional materials comprised of subwavelength inclusions, so-called meta-atoms. By engineering the meta-atoms, one can provide a broad range of functionalities~\cite{Glybovski:2016}. While most metasurface designs target one specific and non-tunable functionality (e.g.~perfect absorption for a specific polarization, wavelength and incidence angle), for many applications it is highly desirable to have tunability and multifunctional features. This can be achieved by introducing globally-tunable materials such as voltage sensitive graphene or locally-tunable elements such as a diode \cite{Liu:2018}. For example, with diodes, programmatically controlled locally-tunable metasurfaces with discrete control of the reflection phase have been demonstrated \cite{Cui:2014,Yang:2016}. Going one step further, achieving continuous control of both reactive and absorptive properties of each meta-atom through an interconnected network of smart controllers can lead to intelligent metasurfaces, implementing a wide range of reconfigurable, diverse functions \cite{Liaskos:2015}. The properties of reflective metasurfaces can be described by the effective surface impedance, which consists of the reactive part and the resistive part. Therefore, the realization of the intelligent metasurfaces requires the local control of both reactance and resistance of the metasurface. In this work, we discuss two distinct designs operating at two different frequency regimes in reflection, with the use of $RC$ loads and graphene. The performance for tunable perfect absorption applications is thoroughly assessed and the practical challenges are discussed.

\section{Switch-Fabric Design for GHz Frequencies}

\begin{figure}
	\centering
	\includegraphics[width=0.86\linewidth]{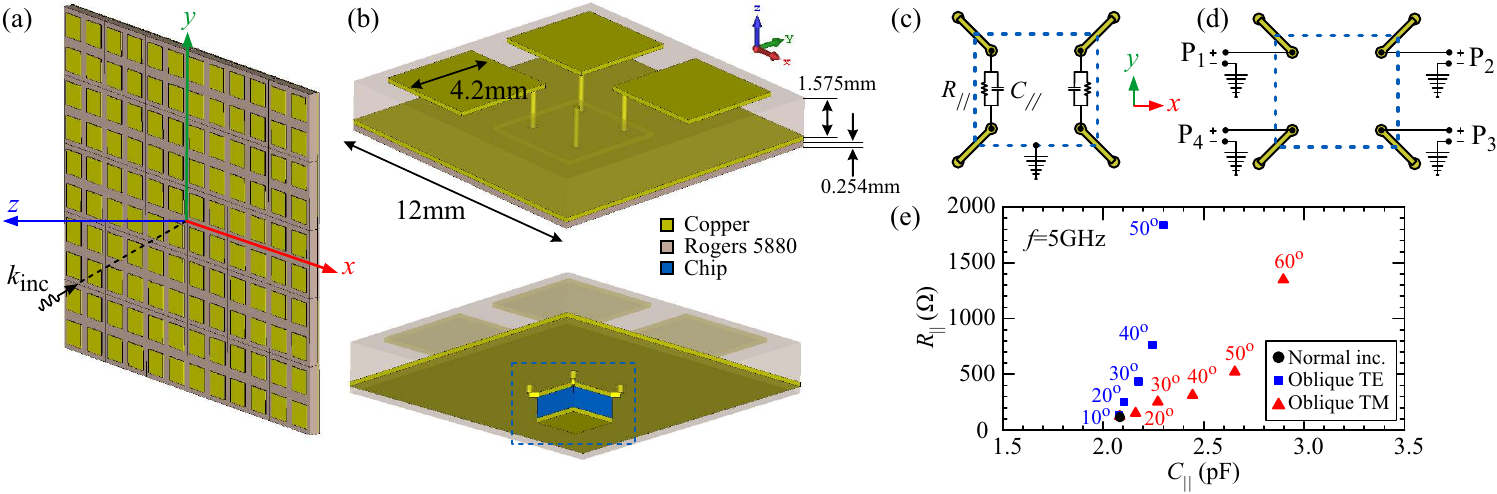}
	\caption{(a)~Switch-fabric-based metasurface design with an impinging wave. (b)~Unit cell comprised of four patches interconnected with a chip by vias through the ground. (c)~Simplified circuit description of the chip supplying variable $R_{\parallel}$ and $C_{\parallel}$. (d)~For an accurate description we extract $S$-parameters of the actual circuit and employ in the EM simulation by discrete ports. (e)~Required $R_{\parallel}$ and $C_{\parallel}$ values for perfect absorption at different angles.} \label{fig:SwitchFabric}
\end{figure}

The simplest way of controlling both the reactance and resistance is to include two lumped elements $R$ (resistor) and $C$ (capacitor) in the meta-atoms. To achieve the selective tuning of $R$ and $C$, we use customized controller chips in which $R$ and $C$ are controlled with two control voltages. The proposed switch-fabric design of the intelligent metasurface is shown in Fig.~\ref{fig:SwitchFabric}(a). The meta-atom, as shown in Fig.~\ref{fig:SwitchFabric}(b), consists of copper patches on a metal-backed dielectric substrate ($\E_r=2.2, \tan\delta=9\times 10^{-4}$ at 5~GHz), connecting with the controller chip through vertical vias. In order to save resources, each chip addresses four patches forming a composite unit-cell.

An important design consideration is the placement of the controller chip below the metallic backplane so that the electromagnetic waves interacting with the metasurface are left unaffected. Furthermore, in this way we do not have to adopt an exact physical description of the intricate, actual integrated circuit implementing the tunable $R$ and $C$ elements. As a result, we can separate the design of the electronics from the electromagnetic design of the unit cell. This is shown in Fig.~\ref{fig:SwitchFabric}(c) where the physical description of the chip is reduced to a parallel connection of lumped $R_{\parallel}$ and $C_{\parallel}$ elements. In this case, configuring the metasurface for a specific function amounts to specifying the $R_{\parallel}$ and $C_{\parallel}$ values uniformly or independently, depending on the desired functionality.

As an example, in Fig.~\ref{fig:SwitchFabric}(e) we specify the required $R_{\parallel}$ and $C_{\parallel}$ values for achieving perfect absorption (when the amplitude of the reflection coefficient $|r|<-30$~dB) at the frequency of $5$~GHz. This lucidly demonstrates the ability to suppress reflection and perfectly absorb the incident wave for a broad range of incidence angles and both TE ($xz$ incidence plane, $\mathbf{E}=E_y\hat{\mathbf{y}}$) and TM ($yz$ incidence plane, $\mathbf{H}=H_x\hat{\mathbf{x}}$) polarizations by properly tuning the chip complex input impedance. Note that for perfect absorption all unit-cells are configured identically; however, our design retains the ability to independently configure each unit-cell which is necessary for other applications, e.g., anomalous reflection, wavefront shaping, etc.


In practice, the $R_{\parallel}$ and $C_{\parallel}$ elements can be implemented by applying appropriate bias voltages to field effect transistors in appropriate topologies (e.g., a varistor can be implemented by a MOSFET in common gate configuration). In order to adopt an accurate circuit description taking actual frequency dependance and parasitics into account, we can extract the circuit $S$-parameters (scattering matrix) and feed them into the electromagnetic full-wave simulation tools via discrete ports, as shown in Fig.~\ref{fig:SwitchFabric}(d). The described procedure enables an accurate co-simulation of electronics and electromagnetics, guaranteeing realistic simulation results.


\section{Graphene-based Design for Terahertz Frequencies}

Another option for fabricating the intelligent metasurface is graphene, due to its gate-tunable electrical conductivity. However, the performance and tunability is restricted by the low achievable mobility of graphene in fabrication. This is due to the unavoidable charge impurities introduced by the substrate during the graphene transfer process, especially for CVD-grown graphene. The low quality of graphene results in its weak light-matter interaction. From the circuit perspective,  low-quality graphene shows high effective shunt resistance (tens of thousands ohm per square) which is much larger than the characteristic impedance of surrounding space \cite{Wang:2017}. This severe impedance mismatch makes the graphene almost transparent for the incident waves (low absorption in graphene).

\begin{figure}[h!]
	\centering
	\subfigure[]{\includegraphics[width=0.34\linewidth]{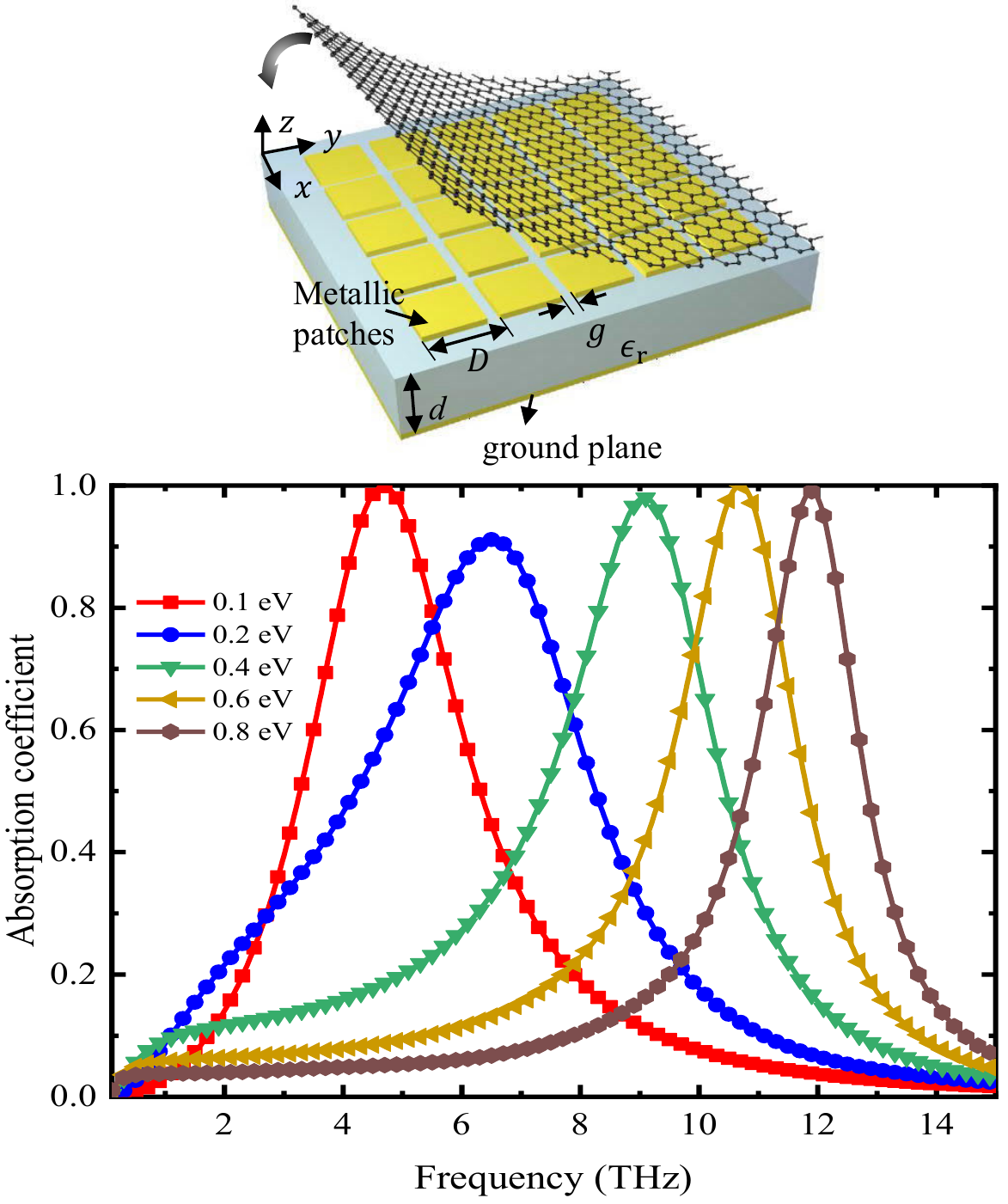}\label{fig:Frequency_tunable_simulation}}\hspace{8mm}
	\subfigure[]{\includegraphics[width=0.35\linewidth]{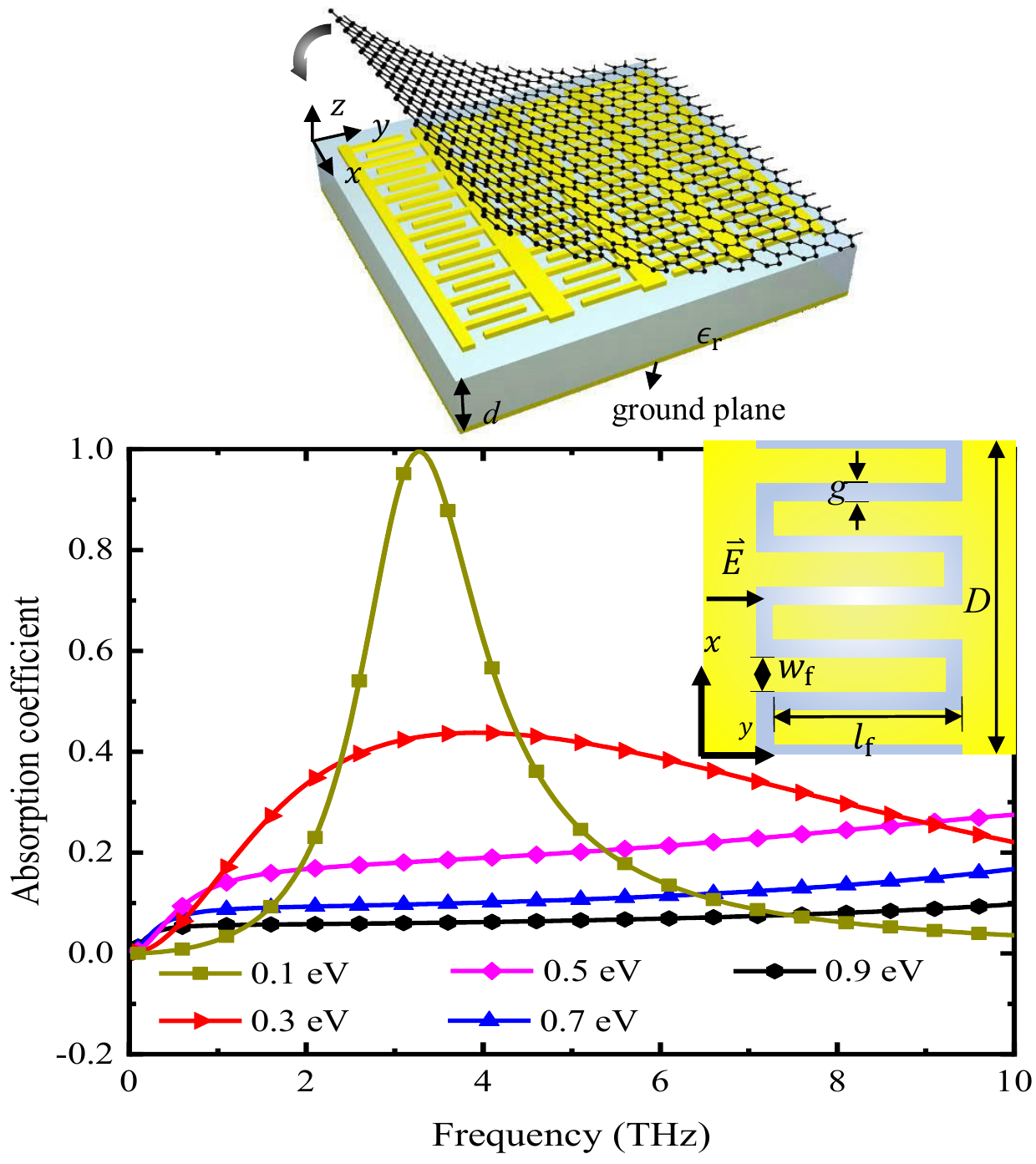}\label{fig:Magnitude_tunable_simulation}}
	\caption{Absorption spectra at different Fermi levels for (a)~tunable perfect absorption with high quality graphene ($\mu=1700~{\rm cm}^{2}{\rm V}^{-1}{\rm s}^{-1}$), (b)~switchable absorber with poor graphene ($\mu=300~{\rm cm}^{2}{\rm V}^{-1}{\rm s}^{-1}$). More details in \cite{Wang:2017}.}\label{proposed_configuration}
 \end{figure}

In order to achieve effective tuning, we introduce a metallic metasurface substrate below a continuous graphene layer, as shown in Fig.~\ref{proposed_configuration}~\cite{Wang:2017}. In this way the high sheet resistance of graphene reduces to near the free space impedance ($\eta_0=$377 $\Omega$). On the other hand, it also acts as a high impedance surface which is necessary for the Salisbury-type absorber. Then, the performance can be tuned by changing the Fermi level. Here, instead of changing the reactance and the resistance at the same time, we demonstrate the individual modification as the first step, with tunable absorption functionalities.


Interestingly, the tunability of absorption depends largely on the mobility of graphene. For example, when the mobility is $\mu\approx1700~{\rm cm}^{2}{\rm V}^{-1}{\rm s}^{-1}$, the frequency for total absorption can be tuned in a wide range (from 5 THz to 12 THz), as shown in Fig. \ref{fig:Frequency_tunable_simulation}.
However, if the carrier mobility is very low,  at the resonant frequency, the effective resistance of graphene changes significantly when tuning the Fermi level of graphene. This rapid and controllable impedance mismatch makes it possible to create a highly efficient switchable absorber.
As an example, we assume a very low mobility value for graphene ($\mu=300~{\rm cm}^{2}{\rm V}^{-1}{\rm s}^{-1}$). Meandered gaps are introduced for reducing the large sheet resistance of such low-quality graphene, as shown in Fig. \ref{fig:Magnitude_tunable_simulation}. When tuning the Fermi level from 0.1 eV to 0.9 eV, the device switches from a perfect absorber to a reflector at 3.2 THz.

\section{Conclusion}
We are targeting to design and fabricate intelligent metasurfaces for electromagnetic wave applications. Due to the tunablility and multifunctionality capabilities, our intelligent metasurfaces can be exploited for a multitude of applications, such as smart antennas for wireless communication, EM energy harvesting, or holography. Furthermore, the concept of intelligent metasurfaces can be also extended to other wave applications, such as in acoustic and elastic waves.

\acknowledgement
This work has received funding from the European Union via the Horizon 2020: Future Emerging Technologies call (FETOPEN-RIA), under grant agreement no. 736876, project VISORSURF.



{\small

\begin{thebibliography}{10}
\setlength{\itemsep}{-1ex}


\bibitem{Glybovski:2016}
S.~B.~Glybovski, S.~A.~Tretyakov, P.~A.~Belov, Y.~S.~Kivshar and C.~R.~Simovski, ``Metasurfaces: From microwaves to visible,'' {\itshape Phys. Rep.,} vol. 634, pp. 1-72, 2016.

\bibitem{Liu:2018}
F.~Liu, A.~Pitilakis, M.~S.~Mirmoosa, \emph{et al.}, ``Programmable metasurfaces: State of the art and prospects,'' {\itshape arXiv preprint}: 1803.04252, 2018.

\bibitem{Cui:2014}
T.~J. Cui, M.~Q. Qi, X.~Wan, J.~Zhao and Q. Cheng, ``Coding metamaterials,
digital metamaterials and programming metamaterials,'' {\itshape Light Sci. Appl.}, vol.~3, pp. 1-9, 2014.

\bibitem{Yang:2016}
H.~Yang, X.~Cao, F.~Yang, J.~Gao, S. Xu, M. Li, X. Chen, Y. Zhao, Y. Zheng and S. Li, ``A programmable metasurface with dynamic polarization, scattering and focusing control,'' {\itshape Sci. Rep.}, vol.~6, p.~35692, 2016.

\bibitem{Liaskos:2015}\textsf{}
C.~Liaskos, A.~Tsioliaridou, A.~Pitsillides, \emph{et al.}, ``Design and development of software defined metamaterials for nanonetworks,'' {\itshape {IEEE} Circuits Syst. Mag.}, vol.~15, no.~4, pp. 12--25, 2015.

\bibitem{Wang:2017}
X.~Wang and S. Tretyakov, ``Tunable perfect absorption in continuous graphene sheets on metasurface substrates,'' {\itshape arXiv preprint}: 1712.01708, 2017.

\end{thebibliography}

}
\end{document}